\documentclass[aps,showpacs,twocolumn,floatfix,amsmath,amssymb,nofootinbib]{revtex4-1}
\usepackage{amssymb}
\usepackage{amsmath}
\usepackage{epsfig}
\usepackage{graphicx}
\usepackage{latexsym}
\usepackage{bm,latexsym}
\usepackage{mathrsfs}
\usepackage{float}

\def\Journal#1#2#3#4{{#1} {\bf #2}, #3 (#4)}

\def\AJ{{Astron. J.}}

\def\APJ{{Astrophys. J.}}

\def\CQG{{Class. Quant. Grav.}}

\def\GRG{{Gen. Relativ. Gravit.}}
\def\IJMPD{{Int. Jour. Mod. Phys. D}}

\def\LRR{{Living Rev. Rel.}}

\def\NATL{{Nature (London)}}

\def\NPBPS{Nucl. Phys. B (Proc. Suppl.)}
\def\PLB{{Phys. Lett.}  B}

\def\PRL{Phys. Rev. Lett.}

\def\PRD{{Phys. Rev.} D}
\def\PR{{Phys. Rep.}}

\def\RMP{{Rev. Mod. Phys.}}

\begin{document}


\title{About matter and dark-energy domination eras in $R^n$ gravity or lack thereof}

\author{Luisa G. Jaime$^{1,2}$}
\email{luisa@nucleares.unam.mx}

\author{Leonardo Pati\~no$^2$}
\email{leopj@ciencias.unam.mx}

\author{Marcelo Salgado$^1$}
\email{marcelo@nucleares.unam.mx}

\affiliation{$^1$Instituto de Ciencias Nucleares, Universidad Nacional
Aut\'onoma de M\'exico, A.P. 70-543, M\'exico D.F. 04510, M\'exico \\
$^2$ Facultad de Ciencias, Universidad Nacional
Aut\'onoma de M\'exico, A.P. 50-542, M\'exico D.F. 04510, M\'exico }


\date{\today}

    
\begin{abstract}
We provide further numerical evidence which shows that $R^n$ models 
in $f(R)$ metric gravity whether produces a late time acceleration in the Universe or a matter domination era (usually a transient one) but not both. 
Our results confirm the findings of Amendola {\it et al.}\cite{Amendola2007a,Amendola2007b,Amendola2007c}, but using a different approach that 
avoids the mapping to scalar-tensor theories of gravity, and therefore, dispense us from any discussion or debate about frames (Einstein {\it vs} Jordan) 
which are endemic in this subject. 
This class of models has been used extensively in the literature as an alternative to the dark energy, but should be considered ruled out for being 
inconsistent with observations. Finally, we discuss a caveat in the analysis by Faraoni~\cite{Faraoni2011},                                 
which was used to further constrain these models by using a chameleon mechanism.
\end{abstract}




\maketitle


\section{Introduction}

$f(R)$ theories of gravity are perhaps the most straightforward modification of general relativity (GR), 
providing an extra geometric component which in some particular cases is capable of generating the accelerated expansion of 
the Universe manifested in supernovae Ia~\cite{SNIa}. A large amount of literature has been accumulated in the past ten years about this 
kind of alternative theories of gravity and is beyond the scope of the present article to make justice to this vast subject 
(see Refs.~\cite{f(R)} for a thorough review). Although some specific $f(R)$ models have shown to be consistent with certain astronomical 
observations, within the Solar System and also cosmological, not every model has the same success, for instance the $f(R)= \lambda R_n (R/R_n)^n$ 
model, simply referred in the literature as to $R^n$. Recently, Amendola {\it et al.} ~\cite{Amendola2007a,Amendola2007b} performed a detailed 
analysis on the cosmological viability of several classes of $f(R)$ models, including $R^n$. Using a dynamical system approach, they concluded 
that for this latter the usual matter era that precedes the accelerated phase with an scale factor $a(t) \sim t^{2/3}$ is generically replaced 
by an non standard era with $a(t) \sim t^{1/2}$ (c.f. Ref.~\cite{Clifton2005} for a complementary analysis), and in the cases where it is possible to 
achieve a usual matter domination epoch the accelerated expansion is not possible. In any instance, the conclusion was that such a model 
is simply unable to reproduce the observed features of our Universe without the addition of some form of dark energy. 

These results have been, however, the object of a debate concerning two issues: 1) the {\it frames} (Einstein {\it vs} Jordan) used in the 
scalar-tensor (ST) approach to analyze the $R^n$ and other models~\cite{Capozziello2006,Amendola2007c,Capozziello2008}; and 2) the analysis of 
the phase space~\cite{Carloni2005,Carloni2009}.

Since the $R^n$ model has been and keeps being considered in the literature (see a complete list of references in~\cite{Faraoni2011}) it is important 
to settle this question with an independent method an beyond any reasonable doubt. 

In this brief report we reanalyze the cosmological case of the $R^n$ model using a different method and 
spanning a wide range of $n$. Our technique does not involve what is usually called the {\it scalar-tensor approach} (ST) where 
a scalar field $\phi = f_R$ is defined in order to map $f(R)$ theories to a Brans--Dicke like theory with $\omega=0$ and a potential. 
Instead, we promote the Ricci scalar itself as an independent degree of freedom~\cite{Jaime2011,Jaime2012a} and in this way we 
circumvent the potential drawbacks associated with the ST approach (e.g. multivalued scalar-field potentials), and in addition 
avoid the long standing issue about frames (Jordan {\it vs} Einstein) which plagues not only the ST method, but also the analysis of scalar-tensor theories 
themselves, and which gave rise precisely to the unnecessary debate mentioned above about the cosmological viability of $R^n$ 
and other class of $f(R)$ theories. As we will show, our approach leads to a rather ``friendly'' system of equations 
which are much more simple to treat than other systems found in the literature and that can be easily solved numerically. 
We had used this method before in the analysis of compact objects~\cite{Jaime2011} and more recently 
in cosmology using different $f(R)$ models~\cite{Jaime2012a,Jaime2012b}. For the cosmological 
analysis at hand, we shall consider the same tools developed in~\cite{Jaime2012a} and adapt them to the case $R^n$.

Our analysis supports the general conclusions of~\cite{Amendola2007a,Amendola2007b} and~\cite{Amendola2007c} (although we do no commit ourselves in assessing the 
soundness of their phase--space analysis) providing a second, independent, strong and 
unambiguous piece of evidence showing that the specific $R^n$ model is not cosmologically viable. In the next section we discuss in detail our 
findings that lead to such conclusion, and we also argue that the analysis put 
forward by Faraoni~\cite{Faraoni2011} to constrain these kind of model 
in the light of the Solar System tests using a chameleon mechanism, is ill founded and requires a deeper review.


\section{$f(R)$ theories}
\label{fR}
The action in $f(R)$ gravity is given by:
\begin{equation}  
\label{f(R)}
S[g_{ab},{\mbox{\boldmath{$\psi$}}}] =
\!\! \int \!\! \frac{f(R)}{2\kappa} \sqrt{-g} \: d^4 x 
+ S_{\rm matt}[g_{ab}, {\mbox{\boldmath{$\psi$}}}] \; ,
\end{equation}
where  $\kappa \equiv 8\pi G_0$ (we use units where $c=1$), $f(R)$ is a sufficiently differentiable but otherwise {\it a priori} arbitrary function of 
the Ricci scalar $R$ and ${\mbox{\boldmath{$\psi$}}}$ represents schematically the matter fields. The field equation obtained from Eq.~(\ref{f(R)}) is:
\begin{equation}
\label{fieldeq1}
f_R R_{ab} -\frac{1}{2}fg_{ab} - 
\left(\nabla_a \nabla_b - g_{ab}\Box\right)f_R= \kappa T_{ab}\,\,,
\end{equation}
where $f_R$ indicates $\partial_R f$, $\Box= g^{ab}\nabla_a\nabla_b$ is the covariant D'Alambertian and $T_{ab}$ is the energy-momentum tensor of matter 
associated with the ${\mbox{\boldmath{$\psi$}}}$ fields. From Eq.~(\ref{fieldeq1}) it is straightforward to obtain the following equation and its 
trace~\cite{Jaime2011,Jaime2012a}
\begin{eqnarray}
\label{fieldeq3}
&& G_{ab} = 
\frac{1}{f_R}\Bigl{[} f_{RR} \nabla_a \nabla_b R +
 f_{RRR} (\nabla_aR)(\nabla_b R) \nonumber\\ 
&&  - \frac{g_{ab}}{6}\Big{(} Rf_R+ f + 2\kappa T \Big{)} 
+ \kappa T_{ab} \Bigl{]} \; ,
\end{eqnarray}
\begin{equation}
\label{traceR}
\Box R= \frac{1}{3 f_{RR}}\Big{[}
\kappa T - 3 f_{RRR} (\nabla R)^2 + 2f- Rf_R
\Big{]} \; ,
\end{equation}
where $(\nabla R)^2:= g^{ab}(\nabla_aR)(\nabla_b R)$ and $T:= T^a_{\,\,a}$.
\footnote{We assume in all the article that a subscript $R$ stands 
for $\partial/\partial_R$.} Equations~(\ref{fieldeq3}) and~(\ref{traceR}) are the basic equations we use in order to find the cosmic evolution in the model $R^n$.  
We have employed this system of equations in the past for several applications, and the reader is invited to consult Refs.~\cite{Jaime2011,Jaime2012a} 
for a detail discussion of this approach. Before analyzing the cosmological situation, it is important to make some remarks 
regarding several issues that arise in this particular model but not in other viable $f(R)$ models. In $f(R)$ theories, one usually demands 
the conditions $f_R >0$ and $f_{RR}>0$. The first one is imposed in order to have a positive definite effective gravitational constant 
$G_{\rm eff}:= G_0/f_R$, while the second condition, is considered in order to avoid instabilities around a possible de Sitter 
background~\cite{Dolgov2003}. We would like to elaborate more about this second point.

In ~\cite{Jaime2011,Jaime2012a} we introduced the potential 
$V(R)$ such that $V_R:=(2f-Rf_R)/3$ which was relevant for tracking the possible de Sitter points allowed by the theory and which correspond 
to trivial solutions of Eq.~(\ref{traceR}) in vacuum. These trivial solutions are given by $R=R_1=const.$ such that $V_R(R_1)=0$, assuming $f_{RR}(R_1)\neq 0$,  
where the effective cosmological constant is $\Lambda_{\rm eff}= R_1/4$. This explains qualitatively why $f(R)$ theories having a de Sitter point 
can potentially produce an accelerated expansion when $R\rightarrow R_1$ and $\rho_{\rm matt}\rightarrow 0$ as the Universe evolves. 
Now, if $f_{RR}(R_1)= 0$, one should define instead ${\tilde V}_R= (2f-Rf_R)/(3 f_{RR})$, being that $f_{RR}$ appears in the denominator 
of Eq.~(\ref{traceR}). Since this situation happens generically in the model $R^n$, we shall consider ${\tilde V}_R$ and not $V_R$. 
Related with the stability analysis is the mass of scalar mode around a de Sitter point $R=R_1$: ${\tilde m}^2:= {\tilde V}_{RR}(R_1)= 
(m^2-f_{RRR} {\tilde V}_R/f_{RR})_{R_1}$, where $m^2= [f_R - R f_{RR}]_{R_1}/[3f_{RR}(R_1)]$, and if $f_{RR}(R_1)\neq 0$ then ${\tilde m}^2 \equiv m^2$ 
since ${\tilde V}_R(R_1)=0$. Usually when $f_{RR}(R_1)\neq 0$, $m^2$ is negative if $f_{RR}(R_1)<0$ (assuming $f_R(R_1)>0$ and $R_1>0$), and in that 
case instabilities may develop rapidly in time~\cite{Dolgov2003}. Thus one should consider theories where $f_{RR}(R_1)> 0$~\footnote{The model $f(R)= R-\mu^4/R$ 
has a de Sitter point at $R_1= \mu^2\sqrt{3}$ and the mass is negative: $m^2= -\mu^2\sqrt{3}$.}, 
and in this case $m^2 >0$ if the critical point at $R_1$ is a minimum of ${\tilde V}(R)$ or $V(R)$.

Let us now focus on $f(R)= \lambda R_n (R/R_n)^n$, where $\lambda$ is a dimensionless constant, and 
$R_n$ is a another constant which in general depends on the choice of $n$, and settles the built-in scale. 
In practice $R_n= \alpha_n H_0^2$, where $\alpha_n$ is a dimensionless constant and $H_0$ is the current Hubble parameter. 
One then has $f_{RR}= \lambda n (n-1)(R/R_n)^{n-2} R_n^{-1}$. We shall not consider the case $n=0$ nor $1$ because 
$n=1$ corresponds to general relativity (GR), for which some sort of dark energy or cosmological constant is required in order 
to explain the accelerated expansion, and for $n=0$ the theory ``disappears'' (i.e. it is too simple), so from now on we assume $n\neq 0,1$.
The condition $f_{RR}> 0$ holds in general provided $n>1$ or $n< -1$, assuming in both cases $R>0$, 
and $f_{RR}$ may vanish only at $R=0$ (we call this point $R_0$) or when $R\rightarrow \infty$ ($R_\infty$). We shall not consider $n<0$ because 
then $f_R= \lambda n (R/R_n)^{n-1}$ becomes negative (assuming $R>0$), and the condition $G_{\rm eff}>0$ is violated. The quantity $f_R$ 
also vanishes at $R_0$ or $R_\infty$, depending on $n$. Finally, ${\tilde V}_R= R^2 (2-n)/[3n(n-1)]$, and thus 
${\tilde V} (R)= R^3 (2-n)/[9n(n-1)] + const.$~\footnote{Had we considered the potential $V(R)$ instead of ${\tilde V}(R)$ 
one would obtain $V_R(R)= 0= \lambda R_n (2-n) (R/R_n)^n/3$, which for $n\neq 2$ and positive has $R=0$ as the only stationary solution in vacuum. 
Therefore in practice $V(R)$ and ${\tilde V}(R)$ single out the same location for the extrema $R_0$ and $R_1$ which correspond to the 
stationary (trivial) vacuum solutions of Eq.~(\ref{traceR}) alluded in the main text for the $R^n$ model.} 
For $n=2$, ${\tilde V}(R)= {const.}$, and any 
$R= R_1\neq 0$ can be a de Sitter point, the specific value $R_1$ depends on the initial conditions when integrating the equations. 
Apart from this ``degenerate'' case, ${\tilde V}_R$ vanishes only at $R_0$. Therefore, for $n\neq 2$ the model $R^n$ does not admit 
de Sitter points and would only be able to generate an accelerated era in a rather transient fashion since far in the future the 
matter contribution dilutes and if $R$ reaches some equilibrium point it will only be at $R_0$ which corresponds to $\Lambda_{\rm eff}=0$. 
The mass ${\tilde m}^2={\tilde V}_{RR}(R)= 2R(2-n)/[3n(n-1)]$, which in this case is to be evaluated at $R=R_1$ or $R=R_0$ (i.e. $R=0$) 
vanishes identically for $n=2$, regardless of the value of the de Sitter point 
$R_1$. Notice that ${\tilde m}^2= 2 m^2/n$, where $m$ was defined above. 
On the other hand, for $n\neq 2$ the only critical point of ${\tilde V}(R)\sim R^3 + const.$ is a saddle point at $R=0$ ($R_0$),
where, as mentioned before ${\tilde V}_R(R_0)$ vanishes, and where ${\tilde m}^2$ vanishes as well regardless of the value of $n$ (we assumed $n\neq 0,1$). 

When a de Sitter point $R_1\neq 0$ exists in vacuum $R_1= 4 \Lambda_{\rm eff}= 12 H_{\rm vac}^2 \neq 0$ (c.f. Eq.~(\ref{Hgen}) with 
$\rho_X= \Lambda_{\rm eff}/\kappa$ and in the limit $\rho\rightarrow 0$). However, with $R_0=0$, one is led to $\Lambda_{\rm eff}=0=H_{\rm vac}^2$. 
Faraoni~\cite{Faraoni2011} overlooked this fact an obtained instead 
$m^2=\frac{1}{3}(f_R/f_{RR}-R)_{R_0}= (2-n)R_0/[3(n-1)]$,\footnote{Notice the missing factors of `2´ and $n$ with respect to ${\tilde m}^2$. The difference arises 
because in our definition of ${\tilde m}^2$ 
we did not assume anything about the critical point precisely because $f_{RR}$ might vanish there. Nonetheless, such factors are irrelevant 
for $R_0=0$ since then $m^2\equiv 0 \equiv {\tilde m}^2$.} 
assuming $R_0\neq 0$, and thus concluding $m^2\neq 0$ for $n\neq 2$. As we just argued, this conclusion is incorrect since 
the only ``de Sitter'' point in the $R^n$ model is $R_0=0$, for $n\neq 2$ and therefore $m^2\equiv 0$~
\footnote{In~\cite{Faraoni2011} the range of the scalar mode was denoted by $s(n)$ which is given by $s(n)= 1/m \sim 1/{\tilde m}$, but 
since both $m$ and ${\tilde m}$ are zero at $R_0$ and at $R_1$ for any $n\neq 0,1$ then $s(n)\rightarrow \infty$, contrary to what was 
found in~\cite{Faraoni2011} for $n\neq 2$, where it was assumed that $H_{\rm vac}^2= R_0/12 \neq 0$, denoted by $H_0^2$ in that reference. 
Here $H_0$ is the actual cosmological constant where all forms of matter (ordinary and the ``geometric dark energy'') 
are taken into account, while $H_{\rm vac}$ is the Hubble expansion when the ordinary matter is neglected and when it is evaluated at the 
stationary solution of Eq.~(\ref{traceRt}). So in~\cite{Faraoni2011} no distinction is made between $H_0$ and $H_{\rm vac}$.}. 
The analysis in~\cite{Faraoni2011} relies 
on the fact that $m^2\neq 0$ and requires the latter to be sufficiently large for the chameleon mechanism to ensue, 
in which case the author concluded $n= 1+\delta$ with $0\leq \delta \leq 5\times 10^{-30}$. Again, that analysis would be valid 
if the model had a true de Sitter point at $R_1\neq 0$ for $n\neq 2$. In light of the previous discussion, we see that the analysis 
in~\cite{Faraoni2011} is no longer sustained nor even required since no matter the value of $n$ (with $n\neq 0,1$) the scalar mode is massless.  
In reality, the chameleon requires a ``thin shell'' condition and an effective mass~\cite{Khoury}, both depending on the density of the environment, 
so $m^2$ by its own does not suffice to analyze such mechanism. But, if it were the case, then  the $R^n$ model would be discarded automatically 
even if $\delta$ were within the above interval (with $\delta\neq 0$) since, the scalar mode being massless, one of the Post-Newtonian parameter would be 
$\gamma\sim 1/2$ whose relative difference with $\gamma_{\rm GR}=1$ is more than four orders of magnitude larger that the maximum value admitted by observations 
$|\gamma-1|\lesssim 2.3\times 10^{-5}$~\cite{Bertotti2003}.
\footnote{It is important to stress that the ``weak-field'', linear or Newtonian limits in $f(R)$ theories are usually studied around a maximum or minimum of 
$\tilde V(R)$. The fact that in this case the critical point is a saddle point indicates that a full non linear analysis is required around that 
point and that such limits are to be reconsidered in $R^n$ gravity. Notice that Eq.~(\ref{traceR}) reads explicitly 
$\Box R= \frac{\kappa T R_n (R/R_n)^{2-n} + \lambda (2-n)R^2}{3\lambda n (n-1)} + (2-n)(\nabla R)^2/R$.}

In the next section we perform a numerical analysis of the full cosmological equations and show that within 
the model $R^n$, including the case $n=2$, an adequate matter dominated era followed by a satisfactory accelerated expansion is 
very unlikely or impossible to happen. 
\begin{figure*}[t!]
\begin{center}
\includegraphics[width=5.9cm]{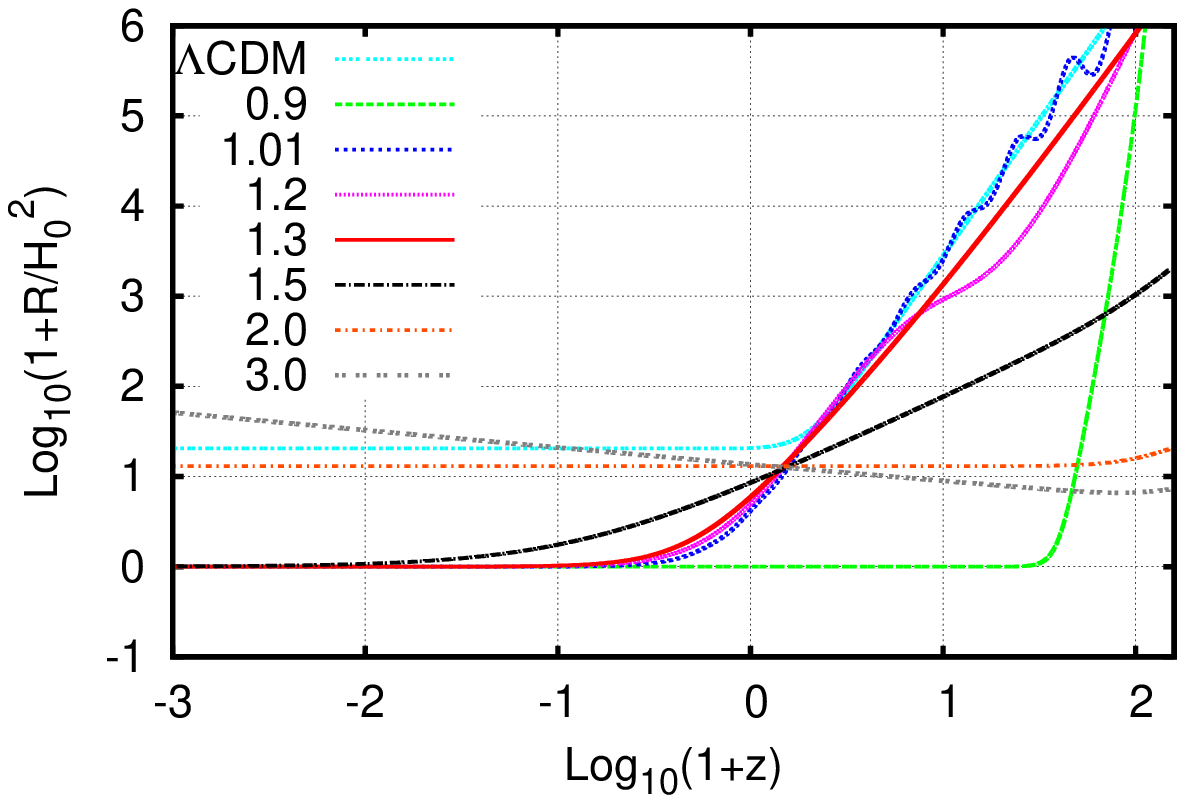} 
\includegraphics[width=5.9cm]{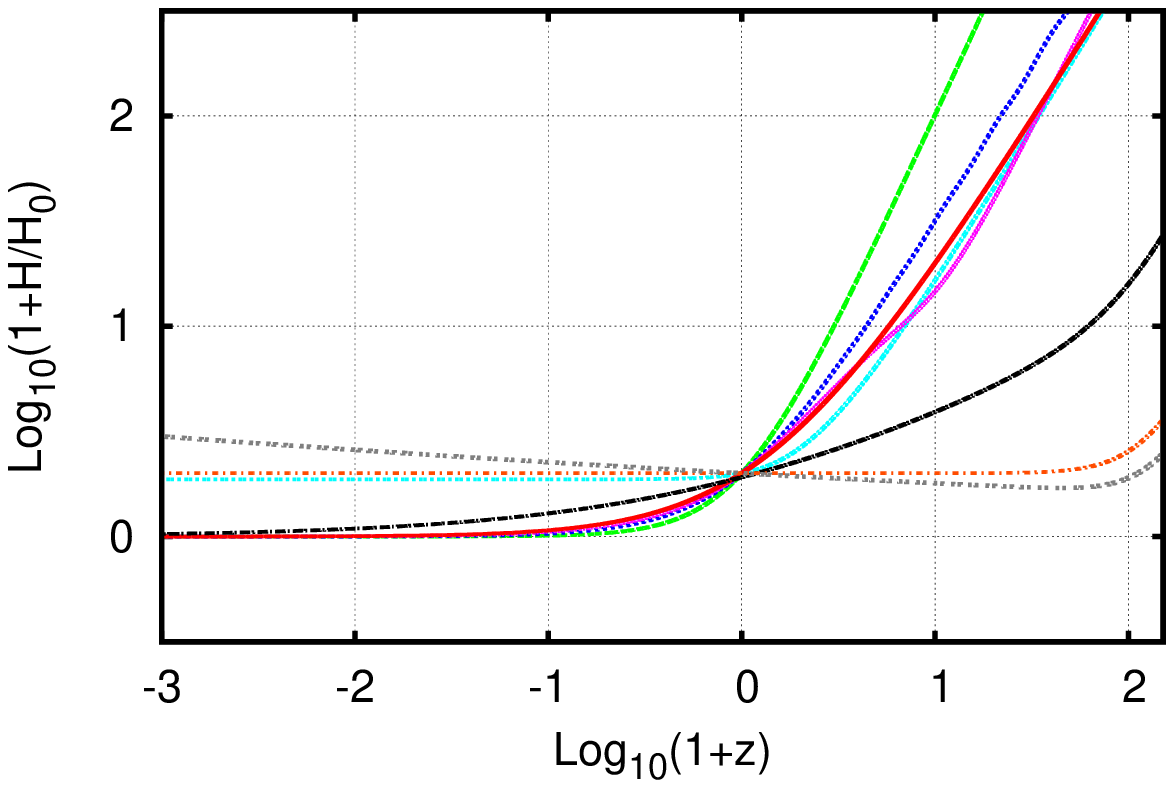}
\includegraphics[width=5.9cm]{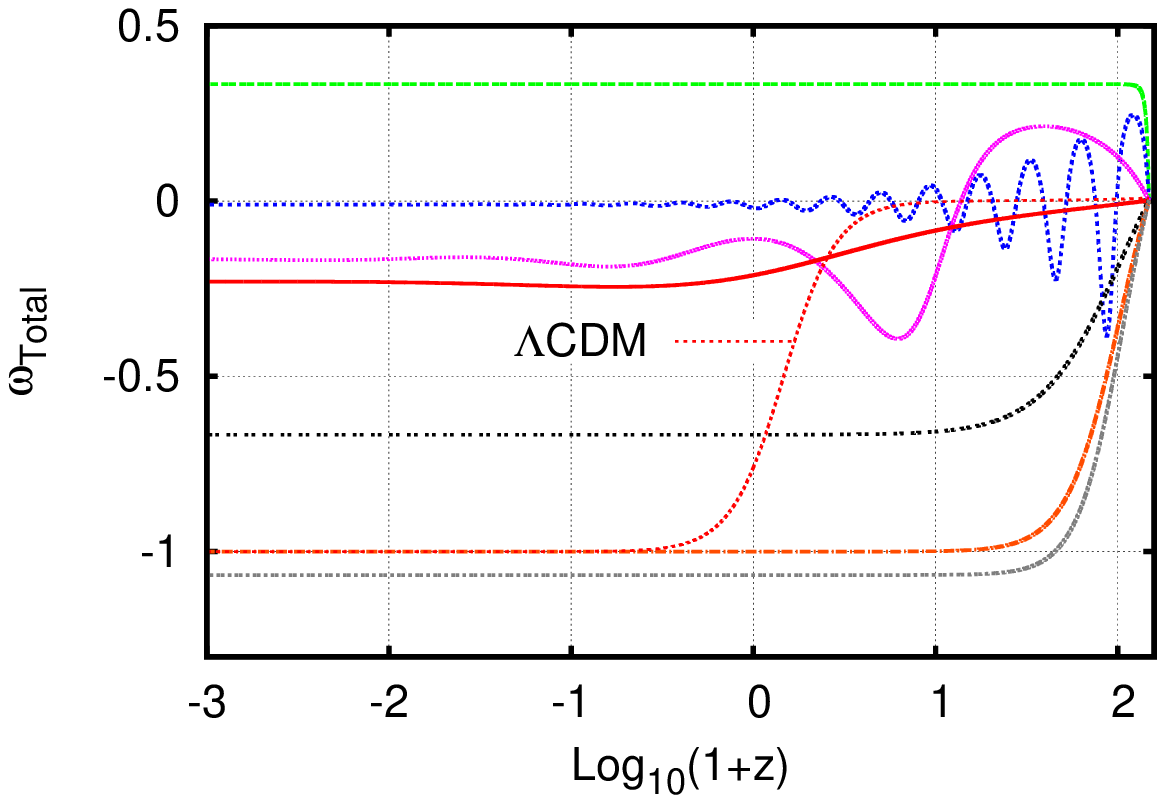}
\end{center}
\vskip -.2cm
\caption{(color online) Ricci scalar (left panel), Hubble parameter (middle panel) and the total EOS $\omega_{\rm tot}$ in $R^n$ gravity 
for several values of the exponent $n$, taking $\lambda=1$ and the constants $\alpha_n= R_n/H_0^2$ as follows: 
$\alpha_{0.9}\sim 577.85$, $\alpha_{1.01}\sim 404.84$, $\alpha_{1.2}\sim 2.02$, $\alpha_{1.3}\sim 1.07$, $\alpha_{1.5}\sim 8\times 10^{-4}$, 
$\alpha_{2}\sim 7.9\times 10^{-6}$, $\alpha_{3}\sim 2.6\times 10^{-6}$. 
The $\Lambda$CDM model is plotted for reference. The plots of the middle and right panels correspond to 
the cases of the left panel. For $n<2$ the Ricci scalar and the Hubble expansion approach zero as $z\rightarrow -1$, while these quantities 
keep growing for $n>2$. The model $n=2$ has an effective cosmological constant and produces $\omega_{\rm tot}=-1$ in the far future. However, 
it does not posses a sufficiently large matter dominated epoch with $\omega_{\rm tot}\sim 0$, as one can appreciate from the right panel. 
None of the models decelerate and accelerate as the $\Lambda$CDM model.}
\label{fig:H-Rn}
\end{figure*}
  
\section{Cosmology in $f(R)$}
We assume a homogeneous and isotropic space-time described by the Friedmann-Robertson-Walker metric:
\begin{equation}
\label{FRWmetric}
ds^2 = - dt^2  + a^2(t) \left[ dr^2 + r^2 \left(d\theta^2 + \sin^2\theta d\varphi^2\right)\right]\,\,\,, 
\end{equation}
where we have taken the flat case $k=0$. From Eqs.~(\ref{fieldeq3}) and ~(\ref{traceR}) we have,
\begin{eqnarray}
\label{traceRt}
&& \ddot R = -3H \dot R - \frac{1}{3 f_{RR}}\left[3f_{RRR} \dot R^2 + 2f- f_R R + \kappa T \right]\!\!,\\
\label{Hgen}
&& H^2 = \frac{\kappa}{3}\left(\rule{0mm}{0.3cm} \rho +\rho_{X}\right) \,\,\,,\\
\label{Hdotgen}
&& \dot{H}= -H^2 -\frac{\kappa}{6}\left\{\rule{0mm}{0.4cm} \rho +\rho_{X}+3\left(p_{\rm rad}+ p_{X}\right) \right\} \,\,\,. 
\end{eqnarray}
where $\dot\,\,= d/dt$ and $H= \dot a/a$, is the Hubble expansion. 
In the above equations we have included the energy 
density $\rho$ associated with matter (baryons and dark matter) and radiation, as well as the geometric dark energy density 
$\rho_{X}$ and pressure $p_{X}$ given explicitly by
\begin{equation}
\label{rhoX}
\rho_{X}=\frac{1}{\kappa f_{R}}\left\{\rule{0mm}{0.5cm} \frac{1}{2}\left( f_{R}R-f\right) -3f_{RR}H\dot{R} + 
\kappa \rho\left(1- f_{R}\right)\right\},
\end{equation}
\begin{equation}
\label{pressX}
p_{X}=-\frac{1}{3\kappa f_{R}}\left\lbrace \frac{1}{2}\left(f_{R}R+f \right) + 3f_{RR}H\dot{R}-\kappa\left(\rho -3 p_{\rm rad} f_R \right)
\right\rbrace \,\,\,.
\end{equation}
These quantities can also be obtained from a covariant and conserved energy-momentum tensor associated with the geometric modifications to GR~\cite{Jaime2012a}.

Notice that the expression for the Ricci scalar 
computed directly from the metric (\ref{FRWmetric}) is given by $R= 6(\dot H + 2 H^2)$ which is, as one can check, compatible 
with the previous evolution equations. Therefore, one can use this latter instead of 
Eq.~(\ref{Hdotgen}). The modified Friedmann Eq.~(\ref{Hgen}) is used only to check the consistency and 
accuracy of our numerical code at every time step and also to fix the initial data (see Ref.~\cite{Jaime2012a} for the details). 
We shall not use $t$ as independent variable but $\alpha= {\rm ln}(a/a_0)$, where $a_0$ is the present value of $a$. 
The corresponding differential equations can be found in~\cite{Jaime2012a}. 

The matter variables obey the conservation equation $\dot \rho_i= -3 H \left(\rho_i + p_i\right)$ for each fluid component labeled by $i$ 
(with $p_{\rm bar,DM}=0$ and $p_{\rm rad}= \rho_{\rm rad}/3$) which integrates straightforwardly and gives rise to the usual expression for 
the energy density of matter plus radiation:
$\rho= (\rho_{\rm bar}^0 + \rho_{\rm DM}^0)(a/a_0)^{-3} + \rho_{\rm rad}^0 (a/a_0)^{-4}$,
where the knotted quantities indicate their values today. The $X$--fluid variables ~(\ref{rhoX}) and (\ref{pressX}) also 
satisfy a conservation equation similar to the one above, but with an equation of state (EOS) $\omega_X:= p_X/\rho_X$ that evolves in cosmic time.

The different domination eras can be tracked via the {\it total} EOS defined by 
$\omega_{\rm tot} = (p_{\rm rad}+p_{X})/(\rho +\rho_{X})$ which using Eqs.~(\ref{rhoX}) and (\ref{pressX}) yields
\begin{equation}
\label{EOSTOT}
\omega_{\rm tot} = -\frac{1}{3}\left[ \frac{\frac{1}{2}\left(f_{R}R+f \right) + 3f_{RR}H\dot{R}-\kappa\rho}
          {\frac{1}{2}\left( f_{R}R-f\right) -3f_{RR}H\dot{R} + \kappa \rho}\right] \,\,\,.
\end{equation}
For instance, during the radiation, matter and geometric-dark-energy dominated eras $\omega_{\rm tot}\sim 1/3,0,-1$ respectively. 
Clearly such values are also correlated with the behavior of the dimensionless densities $\Omega_i= \kappa \rho_i/(3H^2)$ 
which satisfy the constraint $\Omega_{\rm rad} + \Omega_{\rm matt} + \Omega_X =1$ where 
$\Omega_{\rm matt}:=  \Omega_{\rm bar}+  \Omega_{\rm DM}$. The capability of the  $R^n$ model for reproducing the 
correct domination eras will be assessed by the behavior of $\Omega$'s and 
$\omega_{\rm tot}$ during the cosmic evolution relative to the $\Lambda$CDM model. In this regard it is important 
to remark that the $X-$fluid could behave as a matter, radiation or even as a ``ghost'' fluid (one with $\rho_X<0$) depending 
on the value of the exponent $n$, and therefore it could lead to an inadequate evolution history of the Universe. We discuss 
these possibilities in the next section.

\section{Numerical Results and Discussion}

\begin{figure*}[t]
\begin{center}
\includegraphics[width=5.9cm]{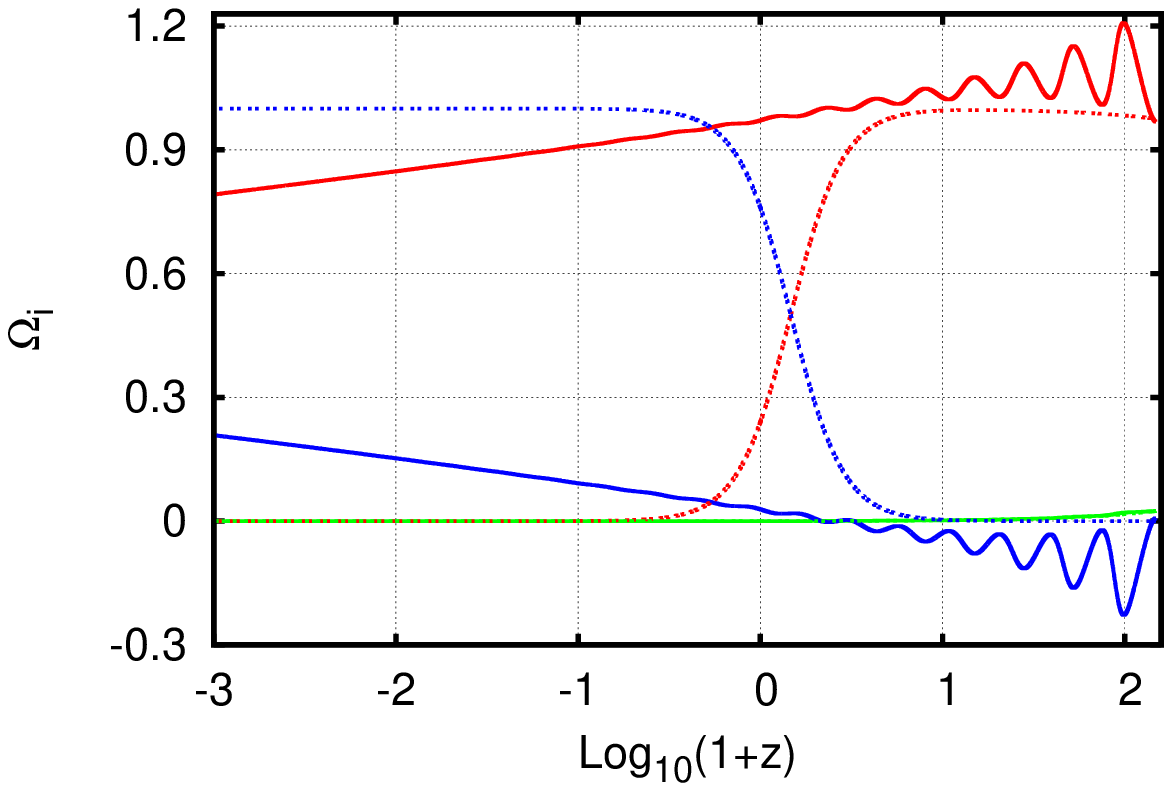}
\includegraphics[width=5.9cm]{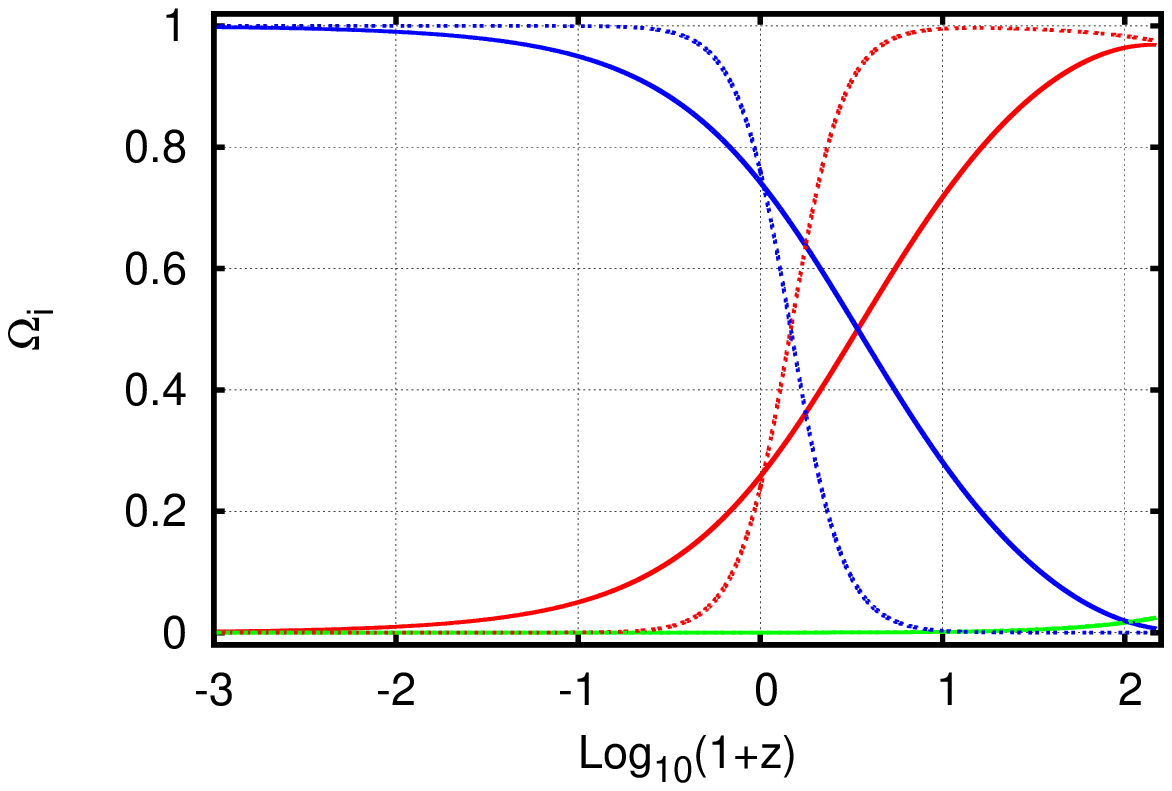}
\includegraphics[width=5.9cm]{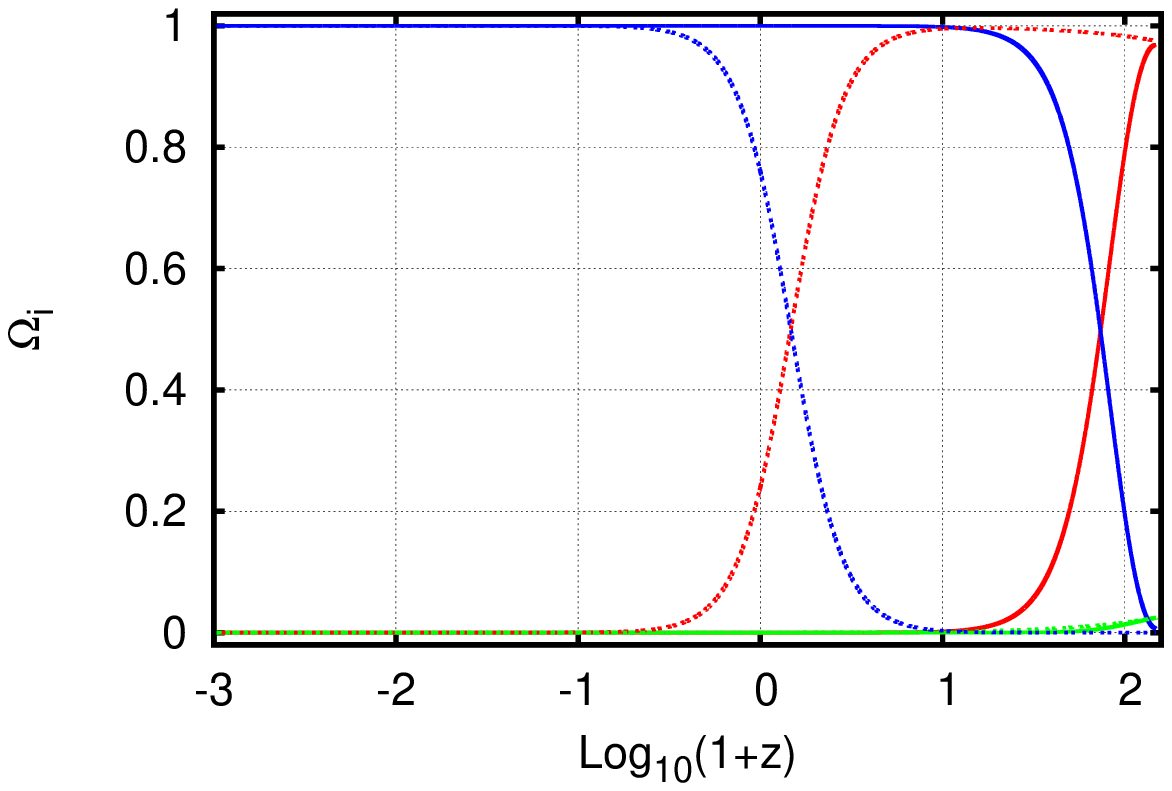}
\end{center}
\vskip -.2cm
 \caption{(color online) Evolution of $\Omega_{\rm matt}$ (red solid line),  $\Omega_X$ (blue solid line) and $\Omega_{\rm rad}$ (green solid line) 
in $R^n$ gravity for $n=1.01$ (left panel), 
$n=1.3$ (middle panel) and $n=2$ (right panel). For reference the corresponding quantities of the $\Lambda$CDM model are included 
in each panel (dashed lines). Notice from the left panel that $\Omega_X$ can be negative and $\Omega_{\rm matt}>1$.}
\label{fig:Ome-n}
\end{figure*}

We integrate the differential equations starting at some redshift $z=a_0/a-1$, say $z \sim 150$, by assuming matter domination for all the $n$'s 
in the $R^n$ model that we analyze. We obtain the initial conditions as described in~\cite{Jaime2012a} and find that varying them in 
several ways it turns out impossible to recover the actual abundances of the different components at present time while having an adequate 
accelerating phase. Here we take $\lambda=1$ but our conclusions do not change by choosing other (positive) values. 
This means that compared to the $\Lambda$CDM model, the Universe expands faster or slower depending on $n$ but it never 
reproduces the correct accelerated expansion and matter domination eras within the same model; it reproduces one or the other in the best of 
scenarios but not both. Figure~\ref{fig:H-Rn} shows the evolution of the Hubble parameter and the Ricci scalar from the past at $z \sim 150$ to the far future 
$z \rightarrow -1$ (the current time corresponding to $z=0$). Notice that for $n=2$ the model admits a de Sitter solution with 
$R\rightarrow R_1\approx 12 H_0^2$ as the Universe evolves towards the present time. Since we have taken into account the matter terms, 
the previous equality does not hold exactly, but approximates very well to the expected value, in agreement with our previous analysis of Sec.~\ref{fR}. 
From Figure~\ref{fig:H-Rn} (right panel) we appreciate that for this $n$, the EOS $\omega_{\rm tot}^{z=0}$ is close to $\omega_{\rm obs}\sim -0.75$, 
which is the required value to explain the current accelerated expansion of the Universe. Nevertheless, the matter epoch is very short as 
compared with the $\Lambda$CDM model. For any other value of $n$, a de Sitter point is never reached, instead $R\rightarrow 0$, and $H\rightarrow 0$ 
for $0< n < 2$ and $R$ and $H$ grows in the future for $n> 2$ (c.f. equation in footnote 6). 

The $\Lambda$CDM model compatible with the supernovae data shows that matter starts dominating for 
$z\gtrsim 0.45$ and dark energy for $z\lesssim 0.45$ which correspond respectively to $\omega_{\rm tot} \gtrsim -0.5$, and 
$\omega_{\rm tot} \lesssim -0.5$ reaching $\omega_{\rm tot}\gtrsim -10^{-2}$ for $z\gtrsim 5$, and $\omega_{\rm tot} \lesssim -0.75$ for $z\lesssim 0$. 
The Universe starts accelerating when $\omega_{\rm tot}<-1/3$ at $z\sim 0.8$.
Figure~\ref{fig:H-Rn} shows that for $n\sim n_c$ with $n_c\approx 1.285$ 
a sufficiently large matter dominated era exists with $|\omega_{\rm tot}|\ll 1$, but approaches the value $\omega_{\rm tot}^{z=0}\approx -0.212$ 
which is incompatible with $\omega_{\rm obs}\sim -0.75$. For $n< n_c$, there is never a matter domination era and 
$\omega_{\rm tot}^{z=0}$ is always far from $\omega_{\rm obs}$, and it can even be positive. In particular, for $n<1$, which we include for illustrative 
purposes as it violates the condition $f_{RR}>0$, the model behaves as radiation dominated with $\omega_{\rm tot}\sim 1/3$ and 
the derivatives $f_R$ and $f_{RR}$ blow up when $R\rightarrow 0$. Finally, for $n> n_c$, there is never a matter 
dominated epoch, but just a transient one with $\omega_{\rm tot}<0$ interpolating monotonically between $\omega_{\rm tot}\sim 0$ and a negative value 
at $z=0$. Among these values, for $n>3$ basically all the models behave identically with $\omega_{\rm tot}\sim -1.067$ 
as $z\rightarrow -1$. Figure~\ref{fig:Ome-n} shows the corresponding evolution of the fractions $\Omega_X$, $\Omega_{\rm matt}$ and $\Omega_{\rm rad}$ for 
a prototype of examples that qualitatively encompasses the rest of the cases, 
and are compared with the $\Lambda$CDM model. For $n\sim n_c$ the abundances are similar to $\Lambda$CDM, particularly 
at the present epoch ($z=0$), but as we mentioned above, the model is unable to accelerate the Universe properly. 
For $n>n_c$ the matter domination epoch is very short (in agreement with the behavior of $\omega_{\rm tot}$ in Fig.~\ref{fig:H-Rn}). 
Finally, for $n<n_c$, 
$\Omega_X$ can even become negative, with $\Omega_{\rm matt}$ possessing phases of superdomination (i.e. $\Omega_{\rm matt}>1$ in those phases) where 
$\omega_{\rm tot}$ can become positive. 
In particular when we take $n = 1+ \epsilon $ with $|\epsilon| \ll 1$, the denominator in Eq.~(\ref{traceRt}), or equivalently in 
Eq.~(\ref{traceR}), becomes very small (c.f. equation in footnote 6) producing 
an important contribution on the r.h.s. of the differential equation for $R$. The cosmological evolution 
for such values of $n$ is then rather different from general relativity. For instance, taking $n=1.01$ we appreciate from 
Fig.~\ref{fig:H-Rn} (right panel) that $\omega_{\rm tot}$ oscillates around a value near zero due to the oscillations of $R$ (left panel). 
This oscillatory behavior can also be appreciated in $\Omega_X$ and $\Omega_{\rm matt}$ from Fig.~\ref{fig:Ome-n} (left panel). The amplitude of these 
oscillations are damped so that $\omega_{\rm tot}\rightarrow 0$ in the present ($z=0$) and future, thus the Universe does not accelerate. On the 
other hand, for $\epsilon<0$, say $\epsilon = -0.1$ ($n=0.9$), $\omega_{\rm tot}\sim 1/3$, as mentioned before, and the Universe behaves as radiation dominated.

Carloni {\it et. al.}~\cite{Carloni2009} following~\cite{Carloni2005}, performed a cosmological analysis 
using a dynamical system approach based on a first order system of equations which is different from ours and 
which was useful for a qualitative description of the cosmological evolution in $R^n$ gravity. 
In their approach they found the fixed points (stable, unstable or saddle) of this and other $f(R)$ models which can represent the 
matter or the accelerated phases in the Universe. As argued by these authors some of these fixed points are different from the ones found 
in~\cite{Amendola2007b} which, as they stressed, might change the conclusions therein. 
Nevertheless, the authors in ~\cite{Carloni2009} acknowledge that their analysis is only qualitative 
as the fixed points might not even be connected, and that an accurate numerical analysis is required. 
This is precisely that we have performed here.

In summary, the homogeneous and isotropic cosmology in $R^n$ gravity seems to show a complete disagreement with what is required to explain the current 
features of the actual Universe. Since our numerical integration 
was performed by including the whole mixture of components in the Universe, even if radiation is relatively small, and without any identification with a 
scalar-tensor theory in any frame whatsoever, the generic problems in the $R^n$ model seem real and are not due to any artifact concerning the ST approach or 
due to any inconsistency regarding a {\it phase space} analysis as objected in~\cite{Capozziello2006,Capozziello2008,Carloni2005,Carloni2009}. 
Thus, we strongly support the conclusion that $R^n$ is not a cosmologically viable candidate, 
unless a curvature $k\neq 0$ changes things dramatically and makes everything fit with observations. But in such occurrence, a non standard 
inflationary paradigm has to be called for explaining the origin of cosmological perturbations.

In~\cite{Jaime2012a,Jaime2012b} we explored other $f(R)$ models that can produce a successful background cosmology 
(i.e. without taking into account perturbations) but needless to say, a detailed scrutiny is required in all possible ambits 
before considering $f(R)$ theories as a serious threat to general relativity.


\acknowledgments
This work was supported in part by DGAPA--UNAM grants IN117012--3, IN115310, IN112210, IN110711 
and SEP--CONACYT 132132. L.G.J. acknowledges support from scholarship CEP--UNAM.



\end{document}